\documentstyle[12pt,epsfig]{article}
\newlength{\dinwidth}
\newlength{\dinmargin}
\begin{document}
%--------------------------------------------------------------------------
\def\bold#1{\setbox0=\hbox{$#1$}%
     \kern-.025em\copy0\kern-\wd0
     \kern.05em\%\baselineskip=18ptemptcopy0\kern-\wd0
     \kern-.025em\raise.0433em\box0 }
\def\slash#1{\setbox0=\hbox{$#1$}#1\hskip-\wd0\dimen0=5pt\advance
%\dimen0 by-\ht0\advance\dimen0 by\dp0\lower0.5\dimen0\hbox
         to\wd0{\hss\sl/\/\hss}}
%--------------------------------------------------------------------------
\newcommand{\be}{\begin{equation}}
\newcommand{\ee}{\end{equation}}
\newcommand{\bea}{\begin{eqnarray}}
\newcommand{\eea}{\end{eqnarray}}
\newcommand{\nn}{\nonumber}
\newcommand{\dd}{\displaystyle}
\newcommand{\bra}[1]{\left\langle #1 \right|}
\newcommand{\ket}[1]{\left| #1 \right\rangle}
\newcommand{\spur}[1]{\not\! #1 \,}
%--------------------------------------------------------------------------
\thispagestyle{empty} 
\vspace*{1cm} 
\rightline{BARI-TH/03-449}
\vspace*{2cm}
\begin{center}
  \begin{LARGE}
  \begin{bf}
Coupling $g_{f_0 K^+ K^-}$ and the structure of $f_0(980)$
\vspace*{0.5cm}
  \end{bf}
  \end{LARGE}
\end{center}
\vspace*{8mm}
\begin{center}
\begin{large}
P. Colangelo and  F. De Fazio
\end{large}
\end{center}
\begin{center}
\begin{it}
Istituto Nazionale di Fisica Nucleare, Sezione di Bari, Italy
\end{it}
\end{center}
\begin{quotation}
\vspace*{1.5cm}
\begin{center}
  \begin{bf}
  Abstract\\
  \end{bf}
  \end{center}
We use light-cone QCD sum rules to evaluate the strong coupling
$g_{f_0 K^+ K^-}$ which enters in several analyses concerning the scalar
$f_0(980)$ meson. The result:  $6.2\le g_{f_0 K^+ K^-}\le 7.8$ GeV is larger
than in previous determinations.

\end{quotation}
\newpage
\baselineskip=18pt
\vspace{2cm}
\noindent
\section{Introduction}
Light scalar mesons are the subject of an intense and continuous scrutiny
aimed at clarifying several aspects of their nature that still need to
be unambiguously established \cite{rassegna,Close:2002zu}.
From the experimental point of view, these particles are difficult
to resolve because of the strong overlap with the continuum background. On the
other hand, the identification is made  problematic since both
quark-antiquark $(q{\bar q})$ and non $q{\bar q}$ scalar states are expected
to exist in the energy regime below 2 GeV. For example, lattice QCD and
QCD sum rule analyses indicate that the lowest lying glueball is a
$0^{++}$ state with mass in the range 1.5-1.7~GeV \cite{Morningstar:1999rf}.
Actually, the observed light scalar states are too numerous to be accomodated
in a single $q \bar q$ multiplet,  and therefore it has been
suggested  that some of them  escape the quark model interpretation.
In addition to glueballs,  other interpretations include
multiquark states and admixtures of quarks and gluons.

Particularly debated  is the nature of the meson $f_0(980)$. Among
the oldest suggestions, there is the proposal that quark
confinement could be explained through the existence of a state
with vacuum quantum numbers and mass close to the proton mass
\cite{Close:1993ti}. On the other hand, following the quark
model and considering the strong coupling to kaons, $f_0(980)$
could be interpreted as an $s{\overline s}$ state
\cite{tornqvist,Tornqvist:1995kr,roos,scadron}.
However, this does
not explain the mass degeneracy between $f_0(980)$ and $a_0(980)$
interpreted as a $({u{\overline u} -d {\overline d})/\sqrt{2}}$
state. A four quark $qq{\overline{qq}}$ state interpretation has
also been proposed \cite{jaffe}. In this case, $f_0(980)$ could
either be nucleon-like \cite{ivan}, {\it i.e.} a bound state of
quarks  with symbolic quark structure $f_0={s{\overline s}({ u
{\overline u}+d {\overline d})/ \sqrt{2}}}$, the $a_0(980)$ being
$a_0=s {\overline s}( u {\overline u}-d {\overline d}) /
\sqrt{2}$, or deuteron-like, {\it i.e.} a bound state of hadrons.
If $f_0$ is a bound state of hadrons, it is usually referred to as
a $K {\overline K}$ molecule \cite{isgur,closebook,kaminski,shev}.
In the former of these two possibilities the mesons are treated as
point-like, while in the latter they should be considered
as extended objects.

The identification of the $f_0$ and of the other lightest
scalar mesons with the Higgs
nonet of a hidden U(3) symmetry has also been suggested 
\cite{Tornqvist:2002bx}.
Finally, a different
interpretation consists in considering  $f_0(980)$  as the result
of a process in which strong interaction enriches a pure ${\bar
q}q$ state with other components, such as $\ket{K{\bar K}}$,
a process known as hadronic dressing
\cite{Tornqvist:1995kr,Boglione:1996uz}; such an interpretation is
supported in
\cite{Close:2002zu,tornqvist,Tornqvist:1995kr,scadron,Tornqvist:2000ju,Shakin:gn,DeFazio:2001uc}.
In ref. \cite{Boglione:2002vv} it has been shown that
the experimentally observed lightest scalar
particles in the I=1 and I=1/2 sectors can be reproduced in this way,
starting from a bare  $q {\bar q}$ and $s{\bar q}$ structure respectively
($q$ being a light non strange quark). On the other hand, I=0 states are
the most elusive ones, since  there are two possible
bare structures, $q {\bar q}$ and $s {\bar s}$, which could
not only undergo hadronic dressing, but also mix through hadronic
loops. The resulting picture strongly depends on the couplings of
the bare structures to the hadronic channels.

Several experimental analyses aimed at discriminating among the
different possibilities. In particular,
the radiative $\phi \to f_0 \gamma$ decay
mode has been identified as  an effective tool for such a purpose
\cite{ivan,closebook,Close:2001ay}. As a matter of fact, if $f_0$
has a pure strangeness component $f_0=s{\bar s}$, the dominant
$\phi \to f_0 \gamma$ decay mechanism is the direct transition,
while  in the four-quark scenario 
$\phi \to f_0 \gamma$ is expected to proceed through kaon loops 
with a branching fraction depending on the specific bound state structure
\cite{closebook,Close:2001ay}.

An important hadronic parameter entering in several analyses involving
$f_0(980)$ is the strong coupling $g_{f_0 K^+ K^-}$. Indeed,
the kaon loop diagrams contributing to $\phi \to f_0 \gamma$ are expressed in
terms of  $g_{f_0 K^+ K^-}$, as well as in terms of the coupling
$g_{\phi K^+ K^-}$ which can be inferred from experimental data
on $\phi$ meson decays. The coupling $g_{f_0 K^+ K^-}$ can
be obtained from various processes, and
we shall present an overview of the determinations in the last part of
this paper. It is interesting to carry out a calculation
in a framework based on QCD, trying to point out
what is a distinctive feature of the scalar
particles, i.e.  their large couplings to the hadronic states.

The present study is devoted to a determination of
$g_{f_0 K^+ K^-}$ by light-cone QCD sum rules, a method
applied to the calculation of several hadronic parameters both in the
light, both in the heavy quark sector \cite{lightconesr,Colangelo:2000dp}.
The analysis and the numerical results are presented in Section 2, while
a summary of the experimental data and of other theoretical
determinations  is given in Section 3. 

\section{Coupling $g_{f_0 K^+ K^-}$ by light-cone QCD sum rules}

In order to evaluate the strong coupling $g_{f_0 K^+ K^-}$,
defined by the matrix element:
\be
\langle K^+(q) K^-(p)|f_0(p+q) \rangle=g_{f_0 K^+ K^-}\,\,\, ,
\label{gf0kk}
\ee
we consider the correlation function
\be
T_\mu(p,q)=i \int d^4x \, e^{i p \cdot x} \,
\langle{K^+(q)}|T[J_\mu^K(x) J_{f_0}(0)]|0\rangle \,\,\, .
\label{corr}
\ee
The quark currents $J_\mu^K$ and $J_{f_0}$ represent
the axial-vector $J_\mu^K={\bar u}\gamma_\mu \gamma_5 s$
and the scalar $J_{f_0}={\bar s}s$ current, respectively,
while the external kaon state has four momentum $q$, with $q^2=M_K^2$.
The choice of the  $J_{f_0}={\bar s}s$ current does not imply that
$f_0(980)$ has a pure ${\bar s}s$ structure, but it simply amounts
to assume that  $J_{f_0}$ has a non-vanishing matrix element between
the vacuum and  $f_0$ 
\cite{DeFazio:2001uc,Aliev:2001mm}.
Such a matrix element, as mentioned below, has been 
derived by the same sum rule method.

Exploiting Lorentz invariance,
$T_\mu$ can be written in terms of two independent invariant functions,
$T_1$ and $T_2$:
\be
T_\mu(p,q) = i \, T_1(p^2,(p+q)^2) \, p_\mu + T_2(p^2,(p+q)^2) \, q_\mu \, .
\label{tmu}
\ee
The analysis of the correlation function in
eq.(\ref{corr}),  following the general strategy of QCD sum rules,
allows us to obtain a quantitative estimate of  $g_{f_0 K^+ K^-}$.
The method consists in representing
$T_\mu$ in terms of the contributions of hadrons
(one-particle states and the continuum) having non-vanishing matrix elements
with the vacuum and the currents $J_\mu^K$ and $J_{f_0}$,
and matching such a representation with a QCD expression computed
in a suitable region of the external momenta $p$ and $p+q$
\cite{Craigie:1982ng}.

Let us consider, in particular,  the invariant function $T_1$ in
eq.(\ref{tmu}) that can be represented  by a dispersive formula
in the two variables $p^2$ and $(p+q)^2$:
\be
T_1(p^2,(p+q)^2)= \int ds ds^\prime {\rho^{had}(s,s^\prime) \over
(s-p^2) [s^\prime-(p+q)^2]} \, . \label{disp}
\ee
The hadronic spectral density $\rho^{had}$ gets contribution from
the single-particle  states $K$ and $f_0$,
for which we define  current-particle matrix elements:
\bea
\langle f_0(980)(p+q)|J_{f_0}|0 \rangle = M_{f_0} {\tilde f}
\label{ftilde}\\
\langle 0 |J_\mu^K| K (p)\rangle = i f_K p_\mu \,\,\, ,
\label{fk}
\eea
as well as from higher resonances and a continuum of states
that we assume to contribute in a domain $D$ of the $s,s^\prime$ plane,
starting from two thresholds $s_0$ and $s_0^\prime$.
Therefore, neglecting the  $f_0$  width,
the spectral function $\rho^{had}$ can be modeled as:
\be
\rho^{had}(s,s^\prime)= f_K M_{f_0} {\tilde f} g_{f_0 K^+ K^-}
\delta(s-M^2_K) \delta(s^\prime-M_{f_0}^2)
+\rho^{cont}(s,s^\prime)\theta(s-s_0)
\theta(s^\prime-s_0^\prime)\,, \label{rho}
\ee
where $\rho^{cont}$ includes the contribution of the higher resonances and
of the hadronic continuum. The resulting expression for $T_1$ is:
\be
T_1(p^2,(p+q)^2)= {f_K M_{f_0} {\tilde f} g_{f_0 K^+ K^-} \over
(M^2_K-p^2) (M_{f_0}^2-(p+q)^2)} +
\int_D ds ds^\prime {\rho^{cont}(s,s^\prime) \over
(s-p^2) [s^\prime-(p+q)^2]} \, . \label{disphad}
\ee
We do not consider possible subtraction terms in eq.(\ref{disp})
as they will be removed by a Borel transformation.

For space-like and large external momenta (large $-p^2$,
$-(p+q)^2$) the function $T_1$ can be computed in QCD
as an expansion near the light-cone $x^2=0$. The expansion
involves  matrix elements of non-local quark-gluon operators,
which are defined in terms of kaon distribution amplitudes
of increasing twist. 
\footnote{The short-distance expansion of the
3-point vacuum correlation function of one scalar 
$\bar s s$ and two pseudoscalar $\bar s i \gamma_5 q$ densities 
has been considered in \cite{Narison:1996fm}.
The present calculation mainly differs for the possibility of
incorporating an infinite series of local operators \cite{Colangelo:2000dp}.}
The first few terms in the expansion are retained,
since the higher twist contributions are suppressed by powers
of  $1/(-p^2)$ or $1/(-(p+q)^2)$.
As a result, the following expression for $T_1$ is obtained to twist
four accuracy:
\bea
T_1(p^2,(p+q)^2)&=&f_K \int_0^1 du\Bigg\{ {M_K^2 \over m_s}
\varphi_p (u) {1 \over m_s^2-(p+uq)^2} \nonumber \\
&-&2\left[ m_s g_2(u)+{M_K^2 \over 6 m_s} \varphi_\sigma(u)(p
\cdot
q+u M_K^2) \right] {1 \over [m_s^2-(p+uq)^2]^2} \Bigg\} \nonumber \\
&+&f_{3K} M_K^2 \int_0^1 dv \left(2v+{1 \over 2} \right) \int
{\cal D}\alpha_i \varphi_{3K}(\alpha_i) {1 \over \{
[p+q(\alpha_1+v\alpha_3)]^2-m_s^2 \}^2 } \label{t1qcd}  \nonumber \\
&-&4 f_K m_s M_K^2 \Bigg\{ \int_0^1 dv (v-1) \int d \alpha_3 {\hat
\psi}(\alpha_3){1 \over \{m_s^2-[p+q((v-1)\alpha_3+1)]^2\}^3}
\nonumber \\
&+& \int_0^1 d \alpha_3 \int_0^{1-\alpha_3} d \alpha_1 {\hat
\phi}(\alpha_i) {1 \over \{m_s^2-[p+q(\alpha_1+v \alpha_3)]^2\}^3
} \Bigg\} \; . \label{t1}
\eea
The functions $\varphi_p$ and $\varphi_\sigma$
appearing in eq.(\ref{t1}) are kaon distribution amplitudes
defined by the matrix elements
\begin{eqnarray}
<K(q)| {\bar u} (x) i \gamma_5 s(0) |0> &=& {f_K M_K^2 \over m_s}
\int_0^1 du \; e^{i u q \cdot x} \varphi_p(u)  \hskip 3 pt ,
 \label{pscal}  \\
<K(q)| {\bar u} (x) \sigma_{\mu \nu} \gamma_5 s(0) |0> &=&i(q_\mu
x_\nu-q_\nu x_\mu)  {f_K M_K^2 \over 6 m_s} \int_0^1 du \;
e^{i u q \cdot x} \varphi_\sigma(u)
 \label{psigma}
\end{eqnarray}
$m_s$ being the strange quark mass (we put to zero the mass of the
light quarks).\footnote{\baselineskip 15pt
The path-ordered gauge factor $P \exp i g_s
\int_0^1 dt x^\mu A_\mu(tx)$ is not included in the matrix elements
having chosen  the gauge $x^\mu A_\mu=0$.}
 Moreover, $g_2(u)$ is defined by the matrix element
\begin{eqnarray}
<K(q)| {\bar u} (x) \gamma_\mu \gamma_5 s(0) |0> = &-&i f_K q_\mu
\int_0^1 du \; e^{i u q \cdot x} [\varphi_K(u)+x^2 g_1(u)]  \nonumber \\
&+&f_K (x_\mu - {q_\mu x^2 \over q \cdot x})
\int_0^1 du \; e^{i u q \cdot x} g_2(u)  \,\,\, . \label{g2}
\eea
Kaon matrix elements of quark-gluon operators also contribute to eq.(\ref{t1});
they are parameterized in terms of twist three and
twist four distribution amplitudes:
\begin{eqnarray}
& &<K(q)| {\bar u} (x) \sigma_{\alpha \beta} \gamma_5
g_s G_{\mu \nu}(v x)s(0) |0>=
\nonumber \\
&&i f_{3 K}[(q_\mu q_\alpha g_{\nu \beta}-q_\nu q_\alpha g_{\mu
\beta}) -(q_\mu q_\beta g_{\nu \alpha}-q_\nu q_\beta g_{\mu
\alpha})] \int {\cal D}\alpha_i \; \varphi_{3 K} (\alpha_i)
e^{iq \cdot x(\alpha_1+v \alpha_3)} , \label{p3k}
\end{eqnarray}
\begin{eqnarray}
& &<K(q)| {\bar u} (x) \gamma_{\mu} \gamma_5 g_s G_{\alpha
\beta}(vx)s(0) |0>=
\nonumber \\
&&f_K \Big[ q_{\beta} \Big( g_{\alpha \mu}-{x_{\alpha}q_{\mu}
\over q \cdot x} \Big) -q_{\alpha} \Big( g_{\beta
\mu}-{x_{\beta}q_{\mu} \over q \cdot x} \Big) \Big] \int {\cal{D}}
\alpha_i \varphi_{\bot}(\alpha_i)
e^{iq \cdot x(\alpha_1 +v \alpha_3)}\nonumber \\
&&+f_K {q_{\mu} \over q \cdot x } (q_{\alpha} x_{\beta}-q_{\beta}
x_{\alpha}) \int {\cal{D}} \alpha_i \varphi_{\|} (\alpha_i)
e^{iq \cdot x(\alpha_1 +v \alpha_3)} \hskip 3 pt \label{gi}
\end{eqnarray}
\noindent
and
\begin{eqnarray}
& &<K(q)| {\bar u} (x) \gamma_{\mu}  g_s \tilde G_{\alpha
\beta}(vx)s(0) |0>=
\nonumber \\
&&i f_K \Big[ q_{\beta} \Big( g_{\alpha \mu}-{x_{\alpha}q_{\mu}
\over q \cdot x} \Big) -q_{\alpha} \Big( g_{\beta
\mu}-{x_{\beta}q_{\mu} \over q \cdot x} \Big) \Big] \int {\cal{D}}
\alpha_i \tilde \varphi_{\bot}(\alpha_i)
e^{iq\cdot x(\alpha_1 +v \alpha_3)}\nonumber \\
&&+i f_K {q_{\mu} \over q \cdot x } (q_{\alpha}
x_{\beta}-q_{\beta} x_{\alpha}) \int {\cal{D}} \alpha_i \tilde
\varphi_{\|} (\alpha_i) e^{iq \cdot x(\alpha_1 +v \alpha_3)} \hskip 3 pt
. \label{git}
\end{eqnarray}
\noindent
The operator $\tilde G_{\alpha \beta}$  is the dual of
$G_{\alpha \beta}$: $\tilde G_{\alpha \beta}= {1\over 2}
\epsilon_{\alpha \beta \delta \rho} G^{\delta \rho} $;
${\cal{D}}\alpha_i$ is defined as ${\cal{D}} \alpha_i =d \alpha_1 d \alpha_2
d \alpha_3 \delta(1-\alpha_1 -\alpha_2 -\alpha_3)$. The function
$\varphi_{3 K}$ is twist three, while the distribution amplitudes
in (\ref{gi}) and (\ref{git}) are twist four.
The functions ${\hat\psi}$ and ${\hat\phi}$
appearing in eq.(\ref{t1}) are defined in terms of
$\varphi_{\bot}$, $\varphi_{\|}$, $\tilde \varphi_{\bot}$ and
$\tilde \varphi_{\|}$ as follows:
${\hat\psi}(\alpha_3)=-\int_0^{\alpha_3}dt \int_0^{1-t} d \alpha_1
\Phi(\alpha_1,1-\alpha_1-t,t)$, and
${\hat\phi}(\alpha_i)=-\int_0^{\alpha_1}dt \Phi(t,1-t-\alpha_3,\alpha_3)$,
with
$\Phi=\varphi_\bot+\varphi_\| + {\tilde \varphi}_\bot+{\tilde \varphi}_\|$.

The sum rule for $g_{f_0 K^+ K^-}$ follows from the approximate
equality of eqs.(\ref{disphad}) and (\ref{t1qcd}).
Invoking global quark-hadron duality, the contribution of the
continuum in (\ref{disphad}) can be identified with the QCD contribution
above the thresholds $s_0, s^\prime_0$.
This allows us to isolate the pole contribution in which the coupling appears. 
The matching between the expressions in
(\ref{disphad}) and (\ref{t1qcd}) can be improved
performing two independent Borel transformations
with respect to the variables $-p^2$ and
$-(p+q)^2$. Defining $M_1^2$ and $M_2^2$ as the Borel parameters
associated to the channels $p^2$ and $(p+q)^2$, respectively, and
using the identity:
\be
 {\cal B}_{M_1^2} {\cal B}_{M_2^2}{( \ell -1)! \over
[m_s^2-(p+uq)^2]^\ell}={(M^2)^{2-\ell} \over M_1^2 M_2^2} \exp
\left( -{m_s^2+q^2 u (1-u) \over M^2} \right) \delta(u-u_0)
\label{borel} \ee with $M^2=\displaystyle{M_1^2 M_2^2 \over
M_1^2+M_2^2}$ and $u_0=\displaystyle{M_1^2  \over M_1^2+M_2^2}$,
we get the following expression for the Borel tranformed eq.(\ref{t1qcd}):
\bea
T_1(M_1^2,M_2^2)&=&f_K M_K^2{e^{-{{\hat m}_0^2
\over M^2}} \over M_1^2 M_2^2} \Bigg\{ {M^2 \over m_s}
\left(\varphi_p(u_0) + {1 \over 6} \varphi_\sigma^\prime(u_0)\right)
 -2 {m_s \over M_K^2}g_2(u_0) +
  \nonumber  \\
&+& {f_{3K} \over f_K} \int_0^{u_0} d \alpha_1
\int_{u_0-\alpha_1}^{1-\alpha_1} {d \alpha_3 \over \alpha_3}
\varphi_{3K}(\alpha_1,1-\alpha_1-\alpha_3,\alpha_3)\left(
2{u_0-\alpha_1 \over \alpha_3}-{1 \over 2} \right) \nonumber \\
&+& 2{ m_s \over M^2}(1-u_0) \int_{1-u_0}^1 {d \alpha_3 \over
\alpha_3^2} {\hat \psi}(\alpha_3)  \label{borelqcd}\\
&-&2 { m_s \over M^2}\left[ \int_0^{1-u_0} {d \alpha_3 \over
\alpha_3} \int_{u_0-\alpha_3}^{u_0} d \alpha_1 {\hat
\phi}(\alpha_i)+ \int_{1-u_0}^1 {d \alpha_3 \over \alpha_3}
\int_{u_0-\alpha_3}^{1-\alpha_3} d \alpha_1 {\hat \phi}(\alpha_i)
\right] \Bigg\} \nonumber
\eea
where  ${\hat m}_0^2=m_s^2+u_0(1-u_0)M_K^2$.
Analogously, a double Borel transformation can be carried out
for the hadronic representation eq.(\ref{disphad}):
\be
T_1(M_1^2,M_2^2)= {e^{-{M_K^2 \over M_1^2}} \over M_1^2}
{e^{-{M_{f_0}^2 \over M_2^2}} \over M_2^2} M_{f_0} {\tilde f} f_K
g_{f_0 K^+ K^-}+{1\over M_1^2 M_2^2 }
\int_D ds ds^\prime \rho^{cont}(s,s^\prime)
e^{-{s \over M_1^2}-{s^\prime \over M_2^2}} \label{borelhad} \,.
\ee
As shown by (\ref{borelhad}), the Borel transformation
exponentially suppresses
the contribution of the higher states and of the continuum;
furthermore, it removes possible subtraction terms in
(\ref{disp}) which depend only on $p^2$ or $(p+q)^2$.

The second term in (\ref{borelhad}) represents the continuum contribution.
In order to identify it with part
the  QCD term  (\ref{borelqcd}), a prescription has been proposed in
\cite{Belyaev:1994zk}. It  consists in considering the symmetric points
$M_1^2=M_2^2=2 M^2$ (corresponding to $u_0=1/2$) in the
$(M_1^2,M_2^2)$ plane and performing the continuum subtraction through
the substitution $e^{-{{\hat m}_0^2 \over M^2}} \to e^{-{{\hat
m}_0^2 \over M^2}}-e^{-{s_0 \over M^2}}$ in the leading-twist
term in (\ref{borelqcd}). Such a prescription is not adeguate in our case, 
where the Borel parameters correspond to channels with different mass
scales and should not be constrained to be equal.
Here we can exploit the property of the amplitudes
$\varphi_p(u)$ and $\varphi_\sigma(u)$ of being polynomials in $u$
(or $1-u$): 
\be 
\varphi_p(u)+{1\over
6}\varphi_\sigma^\prime(u)=\sum_{k=0}^N b_k (1-u)^k 
\ee 
in order to compute  their contribution in the duality region $D$. 
As for the other terms in  (\ref{borelqcd}),
they represent a small contribution to the QCD side of the sum rule, 
and therefore the calculation can leave them unaffected. 

The final expression for  $g_{f_0 K^+ K^-}$ reads:
\bea 
g_{f_0 K^+ K^-} &=& {1\over M_{f_0} \tilde f} \,
e^{M_K^2 \over M_1^2} \, e^{M_{f_0}^2 \over M_2^2} \,
e^{-{{\hat m}_0^2 \over M^2}} \nonumber \\
&\Bigg\{& {M^2 M_K^2 \over m_s} \, \, \sum_{k=0}^N b_k ({M^2 \over M_1^2})^k
\Big[1- e^{-A} \sum_{i=0}^k {A^i \over i!}  +
e^{-A}  {M^2 M_K^2\over M_1^2 M_2^2} {A^{k+1} \over (k+1)!}\Big]
-2 m_s  \, g_2(u_0) \nonumber \\
&+&{f_{3K} M_K^2 \over f_K } \int_0^{u_0} d \alpha_1
\int_{u_0-\alpha_1}^{1-\alpha_1} {d \alpha_3 \over \alpha_3}
\varphi_{3K}(\alpha_1,1-\alpha_1-\alpha_3,\alpha_3)
\left(2{u_0-\alpha_1 \over \alpha_3}-{1 \over 2} \right) \nonumber \\
&+& { 2 m_s M_K^2\over M^2} (1-u_0) \int_{1-u_0}^1 {d \alpha_3 \over
\alpha_3^2} {\hat \psi}(\alpha_3) \label{sumrule} \\
&-&2 {m_s M_K^2\over M^2}\left[ \int_0^{1-u_0} {d \alpha_3 \over
\alpha_3} \int_{u_0-\alpha_3}^{u_0} d \alpha_1 {\hat
\phi}(\alpha_i)+ \int_{1-u_0}^1 {d \alpha_3 \over \alpha_3}
\int_{u_0-\alpha_3}^{1-\alpha_3} d \alpha_1 {\hat \phi}(\alpha_i)
\right] \Bigg\} \nonumber \,,
\eea
with $A={s_0-m_s^2 \over M^2}$ and $s_0$ the smallest continuum threshold.
The prescription in \cite{Belyaev:1994zk} is obtained for $M_1^2=M_2^2$,
$i=0$ and neglecting terms of order $M^2_K$. An interesting
feature of  eq.(\ref{sumrule}) is that, changing  $M_1^2$ and $M_2^2$
independently, it is possible to vary the point $u_0$ where the
distribution amplitudes are evaluated and contribute, while in the standard 
approach the final result is essentially related to the value 
of the distribution amplitudes in a selected point.

The main nonperturbative quantities constituting the input
information in the sum rule (\ref{sumrule}) are the kaon
light-cone wave functions. A theoretical framework for their
determination relies on an expansion in terms of matrix elements
of conformal operators \cite{Braun:1989iv}. For the function
$\varphi_p$,  conformal expansion results in the expression \be
\varphi_p(u,\mu)=\sum_{k} a_k^p(\mu) C_k^{1\over 2} (\xi)
\label{phi} \ee with $\xi=2 u -1$, $a_0^p=1$ and $C_k^{\ell}$
the Gegenbauer polynomials. In (\ref{phi}) we have included the
normalization scale dependence of the distribution amplitude
$\varphi_p$, which appears in the multiplicatively renormalizable
coefficients $a_k^p(\mu)$. 
The nonperturbative information is encoded in the coefficients, 
which are peculiar for the various mesons. In the case of 
kaon, the asymmetry between the
strange and nonstrange quark momentum distribution in the meson
can be taken into account by non-vanishing odd-order coefficients
$a_k^p$. Such $SU(3)$ flavour violating effects have not been
investigated so far for distribution amplitudes of twist larger
than two, and we neglect them in the following, with consequences 
that we shall mention below.
As for the even order coeficients, their updated
values are reported in \cite{Belyaev:1994zk,Ball:1998je}:
$a_2^p=30 \eta_3- {5 \over 2} \rho^2$ and $a_4^p=-3 \eta_3
\omega_3 - {27 \over 20} \rho^2 - {81 \over 10} \rho^2 \tilde
a_2$, with $\tilde a_2=0.2$,  $\eta_3=0.015$,  $\omega_3=-3$ 
at the scale $\mu \simeq 1$ GeV. We have taken into account the meson mass
corrections, related to the parameter $\rho^2={m_s^2\over M_K^2}$,
worked out in \cite{Ball:1998je}.

Analogously, the $\varphi_\sigma$ distribution amplitude can be expressed  as
\be
\varphi_\sigma(u,\mu)=6u(1-u)\sum_{k} a_k^\sigma(\mu) C_k^{3\over 2}(\xi)
\ee
with $a_0^\sigma=1$, $a_2^\sigma= 5 \eta_3- {1\over 2} \eta_3 \omega_3
- {7 \over 20} \rho^2 - {3 \over 5} \rho^2 \tilde a_2$.
For $\varphi_{3K}(\alpha_i)$ and for
the other higher twist distribution amplitudes we refer to the expressions
reported in \cite{Belyaev:1994zk,Ball:1998je}. 

In the analysis of eq.(\ref{sumrule}) we use
$m_s(1\,GeV)=0.14$ GeV \cite{ms}, $M_K=0.4937$ GeV, $M_{f_0}=0.980$ GeV, 
$f_K=0.160$ GeV and ${\tilde f}=(0.180 \pm 0.015)$ GeV \cite{DeFazio:2001uc}.
The threshold parameter $s_0$ is varied around the value
$s_0=1.1$ GeV$^2$ fixed from the determination  of $f_K$
using two-point sum rules \cite{fk}.

%***************************************************************************
\begin{figure}[ht]
\begin{center}
\vskip -1cm
\mbox{\epsfig{file=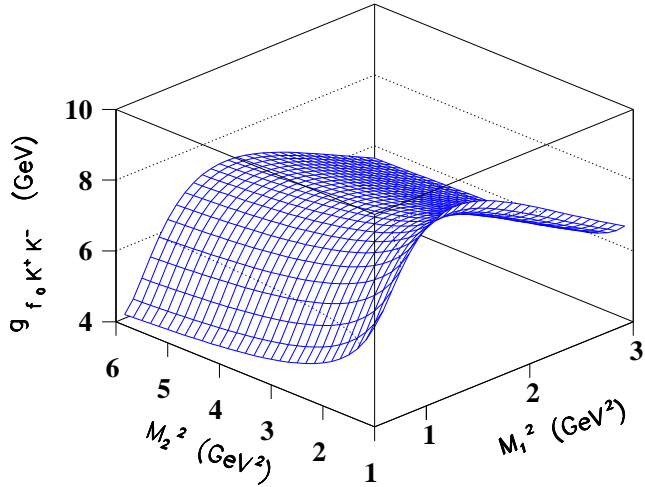, width=10cm}}
\vskip -1cm
\end{center}
\caption{Coupling $g_{f_0 K^+ K^-}$ as a function of the
Borel parameters $M_1^2$ and $M_2^2$, for $s_0=1.1$ GeV$^2$.}
\label{fig:gf0kk_surf}
\end{figure}
%***************************************************************************

The result for $g_{f_0 K^+ K^-}$ versus the Borel parameters
$M_1^2$ and $M_2^2$ is depicted in fig.\ref{fig:gf0kk_surf}. A
stability region where the outcome does not depend on $M_i^2$ can
be selected. Such a region does not correspond to the
line $M_1^2=M_2^2$, but to the range 
$0.8 \le M_1^2 \le 1.6$ GeV$^2$ with $M_2^2$ extending up to $M_2^2\simeq 5$
GeV$^2$. Varying $M_1^2$ and  $M_2^2$ in this region, and changing 
the values of the thresholds and of the other parameters, we obtain the
result  depicted in fig.\ref{fig:gf0kk}, which can be quoted as
$6.2 \le g_{f_0 K^+ K^-}\le 7.8$ GeV.

Let us briefly discuss the uncertainties affecting the numerical result.
As for the $SU(3)_F$ breaking effects rendering the kaon distribution 
amplitudes asymmetric with respect to the middle point, 
the neglect should have a minor role in our approach, due to the possibility 
of exploring wide ranges of the variable $u$ and smoothing the effects of
the actual shapes of the wave functions. Another uncertainty is 
related to the value of
the strange quark mass, $m_s$; since the dependence of the sum rule on 
$m_s$ mainly involves the ratio $M_K^2/m_s$, one can fix this ratio
using chiral perturbation theory, obtaining results in the same
range quoted for $g_{f_0K^+K^-}$.

We can compare now our result with the available
experimental determinations of $g_{f_0K^+K^-}$, as well as with 
the results of other calculations. We shall see how complex the  scenario is.
%
%***************************************************************************
\begin{figure}[ht]
\begin{center}
\vskip -1cm
\mbox{\epsfig{file=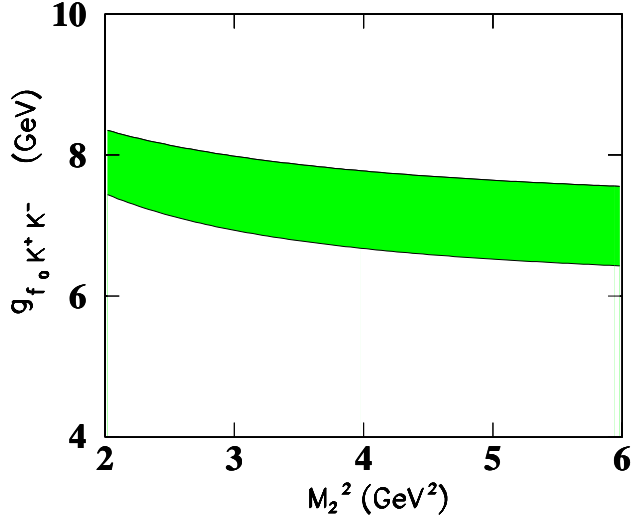, width=10cm}}
\vskip -1cm
\end{center}
\caption{Coupling $g_{f_0 K^+ K^-}$ as a function of the
Borel parameter $M_2^2$, varying $s_0$ in the range
$1.05 \le s_0\le 1.15$ GeV$^2$ and $M_1^2$ in the range
$0.7 \le M_1^2 \le 2.0$ GeV$^2$.} \label{fig:gf0kk}
\end{figure}
%***************************************************************************
%
\section{Other determinations of $g_{f_0K^+K^-}$}

As discussed in the Introduction,
$g_{f_0K^+K^-}$ can be considered in
connection with the radiative $\phi \to f_0 \gamma$ decay mode.
Several analyses go through this decay channel.
KLOE Collaboration at the DA$\Phi$NE collider in Frascati has
examined the decay channel $\phi \to \pi^0 \pi^0 \gamma$
measuring the branching fraction: ${\cal B}(\phi \to \pi^0 \pi^0
\gamma)=(1.09 \pm 0.03_{stat} \pm 0.05_{syst}) \times 10^{-4}$
\cite{Aloisio:2002bt}. The decay mode is supposed to proceed
through $\rho \pi$ intermediate state and through kaon loop
processes, with the kaons annihilating into scalar resonances that
subsequently decay to $\pi^0 \pi^0$. Different fits of the two
pion invariant mass spectrum $\displaystyle{d \Gamma \over d
M_{\pi \pi}}$ are performed in order to measure the parameters of
the scalar states. In a first fit (A) only the contribution of the
intermediate state  $f_0(980)$ is considered, and the three
parameters $M_{f_0},~g^2_{f_0K^+K^-}$ and
$g^2_{f_0K^+K^-}/g^2_{f_0\pi \pi}$ are determined. In a second fit
(B) the  contribution  of a  possible broad scalar $\sigma$ state
is included, and the coupling $g_{\phi \sigma \gamma}$ is
considered as a further parameter. It is  assumed that the two
pion decay modes saturate the $f_0$ width, and that ${\cal B}(f_0
\to \pi^+ \pi^-)=2 ~{\cal B}(f_0 \to \pi^0 \pi^0)$ invoking
isospin symmetry. Fit A provides ${\cal B}(\phi \to f_0
\gamma)=(3.3 \pm 0.2) \times 10^{-4}$ and
$\displaystyle{g^2_{f_0K^+K^-}\over (4 \pi)}=1.29 \pm 0.14~GeV^2$
($\chi^2/{\rm ndf}=109.53/34$). Fit B gives instead: ${\cal
B}(\phi \to f_0 \gamma)=(4.47 \pm 0.21) \times 10^{-4}$ and
$\displaystyle{g^2_{f_0K^+K^-}\over (4 \pi)}=2.79 \pm 0.12~GeV^2$
($\chi^2/{\rm ndf}=43.15/33$). The negative interference between
the contributions of the broad $\sigma$ and the $f_0$ is
responsible of the improvement in the accuracy of the fit. In both
cases sizeable values for $g_{f_0K^+K^-}$ are obtained; they are
reported in Table \ref{tableexp}.

An analogous analysis has been performed by the
CMD-2 Collaboration at the VEPP-2M collider in Novosibirsk.
From a combined fit to the spectra of the decays $\phi \to \pi^+
\pi^- \gamma$ and $\phi \to \pi^0 \pi^0 \gamma$, CMD-2 Collaboration
obtains: ${\cal B}(\phi \to f_0 \gamma)=(2.90 \pm
0.21 \pm 0.65) \times 10^{-4}$ and
$\displaystyle{g^2_{f_0 K^+ K^-} \over (4 \pi)}=1.48 \pm 0.32\,\,GeV^2$
\cite{Akhmetshin:1999di}. A similar result is quoted by the SND
Collaboration at the same VEPP collider:
${\cal B}(\phi \to f_0 \gamma)=(3.5 \pm 0.3
\pm^{1.3}_{0.5}) \times 10^{-4}$ and
$\displaystyle{g^2_{f_0K^+K^-}\over (4 \pi)}=2.47^{+0.73}_{-0.51}~GeV^2$
\cite{Achasov:2000ym}.

Other determinations of $g_{f_0 K^+ K^-}$ rely on the analysis of
different  physical processes. Considering the central $f_0$ production in
$pp$ collisions, the WA102 experiment at CERN gets:
$\displaystyle{g^2_{f_0 K^+ K^-}\over(4 \pi)}=0.38 \pm 0.06 \,\,
GeV^2$ \cite{Barberis:1999cq}. On the other hand, analyzing the
$f_0$ production in $D_s$ decays to three pions, the Collaboration
E791 at Fermilab finds a value compatible with zero
\cite{Gobel:2000es}. These results are also reported in Table
\ref{tableexp}.

In Ref.\cite{ivan} the coupling constant is evaluated for
different values of the phase shift of the elastic background in
the $\pi \pi \to \pi \pi$ reaction, of the ratio $R=g_{f_0
K^+K^-}^2/g_{f_0 \pi^+\pi^-}^2$ and according to different
scenarios for the $f_0$ structure, obtaining  results in a wide range:
$g_{f_0 K^+K^-} \in [1.95,7.3]$ GeV.

The analysis of the decay channel $J/\psi \to \phi K {\overline K}
(\pi \pi)$ has been carried out in Ref.\cite{Escribano:2002iv}.
The $f_0$ pole is described as a Breit-Wigner resonance coupled to
two channels. Two fits of the experimental data are performed
depending upon the $\pi \pi$ phase shift data used, obtaining 
$g_{f_0 K^+K^-}=2.5 \pm 0.15 $ GeV and $g_{f_0 K^+K^-}=2.0
\pm 0.06$ GeV, respectively.

A  prediction for $\displaystyle{g_{f_0 K^+ K^-}}$
based on chiral symmetry and the linear sigma
model,  when no mixing with the $\sigma$ is considered, is:
$g_{f_0 K^+ K^-}=2.24$ GeV \cite{Napsuciale:1998ip}, to be
compared to old determinations $g_{f_0 K^+ K^-}=2.74$ GeV
\cite{goldest}.
 Using the method of the T-matrices,  the value
$g_{f_0 K^+ K^-}=3.8 $ GeV is obtained \cite{Oller:1997ti}.

Considering all the above results one sees that a general consensus on
$g_{f_0 K^+ K^-}$ has not been reached, so far. In particular, experimental
analyses of different processes produce contradicting results. 
The outcome from $\phi \to f_0 \gamma$ points towards sizeable values
of the coupling,  consistent with the
light-cone sum rule result. One has to say that the error quoted
for the experimental determinations, which in general looks small,
mainly accounts for the statistical uncertainties; 
one could infer the size of the
systematical uncertainties comparing different determinations.

As for $g_{f_0 K^+ K^-}$ from $D_s$ decays, presumably
the determination will be  improved at the B factories
by experiments such as   BaBar at SLAC,
thanks to large available samples of $D_s$ mesons. In these measurements
$g_{f_0 K^+ K^-}$ is expected to be determined by coupled channel analyses, 
with $D_s$ decaying to final states containing kaons 
as well as pions \cite{Meadows:2002wu}.

%*******************************************************************
\begin{table}[h]
\caption{Experimental determinations of $g_{f_0 K^+ K^-}$ using different 
physical processes.
Double items refer to two different
fits (see text).} \label{tableexp}
\begin{center}
\begin{tabular}{|| l | c |  c | c ||} \hline \hline
Collaboration &process & $g_{f_0 K^+ K^-}\,(GeV)$ & Ref. \\ \hline
KLOE    &$\phi \to f_0 \gamma \, (A)$&$4.0\pm 0.2 \, (A)$&
\cite{Aloisio:2002bt}\\
        &$\phi \to f_0 \gamma \, (B)$&$5.9\pm 0.1 \, (B)$&\\  \hline
CMD-2   &$\phi \to f_0 \gamma$       &$4.3\pm 0.5$       &
\cite{Akhmetshin:1999di} \\  \hline SND     &$\phi \to f_0 \gamma$
&$5.6\pm 0.8$       & \cite{Achasov:2000ym} \\  \hline \hline
WA102   &$pp$                        &$2.2\pm 0.2$        &
\cite{Barberis:1999cq} \\  \hline \hline E791    & $D_s\to 3 \pi$
&$0.5\pm 0.6$      & \cite{Gobel:2000es}\\ \hline \hline
\end{tabular}
\end{center}
\end{table}
%***********************************************************************
\section{Conclusions}

The purpose of this paper was the evaluation of the strong coupling
constant $g_{f_0 K^+ K^-}$, the value of which is rather
controversial, as it emerges comparing different experimental and theoretical
determinations. In particular, the KLOE Collaboration 
measured a larger value than in  other determinations, with a greater 
accuracy as well. However,  such a result stems from the investigation of
$\phi \to f_0 \gamma$, and therefore it is
mandatory to wait for the study of unrelated processes, namely
the combined analysis of $D_s$ decays to pions and kaons.
The outcome of light-cone QCD sum rules is in keeping
with a large value for the coupling. The uncertainty affecting the
result is intrinsic of the method and does not allow a better comparison
with data. However, the analysis confirms a peculiar aspect of the scalar
states, i.e. their large hadronic couplings,  thus pointing towards a
scenario in which the process of hadronic dressing is favoured. 

\vspace*{0.5cm}
\noindent {\bf Acknowledgments }
We thank A. Palano, T. Aliev and M.E. Boglione for discussions.
We acknowledge partial support from the EC Contract No.
HPRN-CT-2002-00311 (EURIDICE).

\end{document}